\documentclass{emulateapj}

\usepackage{graphicx}
\usepackage{sidecap} 
\usepackage{epsfig}
\usepackage{times}
\usepackage{natbib}
\usepackage{url}
\usepackage{color}
\usepackage{amssymb,amsmath}
\usepackage{ulem} 

 \definecolor{DarkMagenta}{rgb}{0.45,0.0,0.45}  
 
 \definecolor{DarkOrange}{rgb}{0.6,0.4,0}

\newcommand{\hi}{\rm H~\textsc{i}}
\newcommand{\hei}{\rm He~\textsc{i}}
\newcommand{\heii}{\rm He~\textsc{ii}}

\shorttitle{Reconnection driven by CME expansion}
\shortauthors{van Driel-Gesztelyi et al.}

\begin{document}

\title{Coronal magnetic reconnection driven by CME expansion -- the 2011 June 7 event}

\author{
L.~van Driel-Gesztelyi\altaffilmark{1,2,3}, 
D.~Baker\altaffilmark{1}, 
T.~T\"{o}r\"{o}k\altaffilmark{4}, 
E.~Pariat\altaffilmark{2}, 
L.~M. Green \altaffilmark{1}, 
D.~R.~Williams\altaffilmark{1}, 
J.~Carlyle\altaffilmark{1,5}, 
G.~Valori\altaffilmark{2}, 
P.~D\'emoulin\altaffilmark{2},  
B.~Kliem\altaffilmark{1,6,7}, 
D.~M.~Long\altaffilmark{1},  
S.~A.~Matthews\altaffilmark{1},
J.-M.~Malherbe\altaffilmark{2}}
\altaffiltext{1}{University College London, Mullard Space Science Laboratory, Holmbury St Mary, Dorking, Surrey, RH5 6NT, UK}
\altaffiltext{2}{Observatoire de Paris, LESIA, UMR 8109 (CNRS), Meudon-Principal Cedex, France}
\altaffiltext{3}{Konkoly Observatory of the Hungarian Academy of Sciences, Budapest, Hungary}
\altaffiltext{4}{Predictive Science Inc. 9990 Mesa Rim Rd., Ste 170, San Diego, CA 92121, U.S.A.}
\altaffiltext{5}{Max-Planck Institut f\"ur Sonnensystemforschung, 37191 Katlenburg-Lindau, Germany.}
\altaffiltext{6}{Institut f\"ur Physik und Astronomie, Universit\"at Potsdam, Karl-Liebknecht-Str. 24-25, 14476 Potsdam, Germany.}
\altaffiltext{7}{Yunnan Observatory, Chinese Academy of Sciences, Kunming, Yunnan 650011, China}

\begin{abstract}
Coronal mass ejections (CMEs) erupt and expand in a magnetically structured solar corona. Various \textit{indirect} observational pieces of evidence have shown that the magnetic field of CMEs reconnects with surrounding magnetic fields, forming, e.g., dimming regions distant from the CME source regions. Analyzing Solar Dynamics Observatory ({\sl SDO}) observations of the eruption from AR~11226 on 2011 June 7, we present the first direct evidence of coronal magnetic reconnection between the fields of two adjacent ARs during a CME. The observations are presented jointly with a data-constrained numerical simulation, demonstrating the formation/intensification of current sheets along a hyperbolic flux tube (HFT) at the interface between the CME and the neighbouring AR~11227. Reconnection resulted in the formation of new magnetic connections between the erupting magnetic structure from AR~11226 and the neighboring active region AR~11227 about 200 Mm from the eruption site. The onset of reconnection first becomes apparent in the {\sl SDO}/AIA images when filament plasma, originally contained within the erupting flux rope, is re-directed towards remote areas in AR~11227, tracing the change of large-scale magnetic connectivity. The location of the coronal reconnection region becomes bright and directly observable at {\sl SDO}/AIA wavelengths, owing to the presence of  down-flowing cool, dense ($10^{10}$~cm$^{-3}$) filament plasma in its vicinity. The high-density plasma around the reconnection region is heated to coronal temperatures, presumably by slow-mode shocks and Coulomb collisions. These results provide the \textit{first direct observational evidence} that CMEs reconnect with surrounding magnetic structures, leading to a large-scale re-configuration of the coronal magnetic field.         
\end{abstract}


\section{Introduction}

Coronal mass ejections (CMEs) are fundamentally magnetic structures ejected from the Sun. Since magnetic fields are ubiquitous in the corona, CMEs inevitably encounter and interact with other magnetic structures as they expand \citep{Attrill07,vanDriel08}. Such interactions have been successfully modeled via magnetohydrodynamic (MHD) simulations \citep[e.g.,][]{Roussev:2007p23531,Cohen09,Cohen10,Lugaz:2011p23536,Masson13}. The simulations showed that the lateral expansion of CMEs in the low corona could be facilitated by magnetic reconnection between the expanding CME core and the surrounding magnetic environment. \cite{Cohen10}, modeling the 1997 May 12 CME, found that the erupting flux rope field undergoes interchange reconnection with open field lines of the northern coronal hole, and also reconnects with the oppositely oriented overlying magnetic field in the manner of the breakout model \citep{Antiochos99}. The modification of CME connectivity, due to interaction with overlying fields, continued up to the point where the CME reached the outer edge of the streamer belt.  

Observationally, the interaction is revealed by signatures of activity, e.g, non-thermal radio emission \citep[e.g.,][]{Pick98,Maia99,Pohjolainen01,Wen06}, the formation of wide-spread coronal dimming regions \citep[e.g.,][]{Attrill07,Mandrini07,Cohen09}, as well as disturbances (brightenings) along coronal hole boundaries \citep[e.g.,][]{Hudson96, Attrill06, Harra07}. Trans-equatorial loops are also reported to disappear as a consequence of some CMEs \citep[e.g.,][]{Delannee99,Khan00,Pohjolainen01}. 

 However, these are all indirect pieces of evidence of magnetic reconnection between the erupting CME and surrounding magnetic fields, while the actual process and its exact location remain unobserved.  Since the fast expansion of the magnetic flux rope implies a steep decrease of its magnetic field strength, these sporadic reconnection processes involve little magnetic flux at a time, and in the tenuous corona the emission measure is usually too low for any resulting in-situ heating to be imaged. Exceptional circumstances are needed to capture the locations of reconnection regions. Such circumstances were found during a dramatic  filament eruption on 2011 June 7, observed by the {\sl Solar Dynamics Observatory}'s Atmospheric Imaging Assembly ({\sl SDO}/AIA).

\begin{figure}[t]
\centering
\includegraphics[width=1.\linewidth]{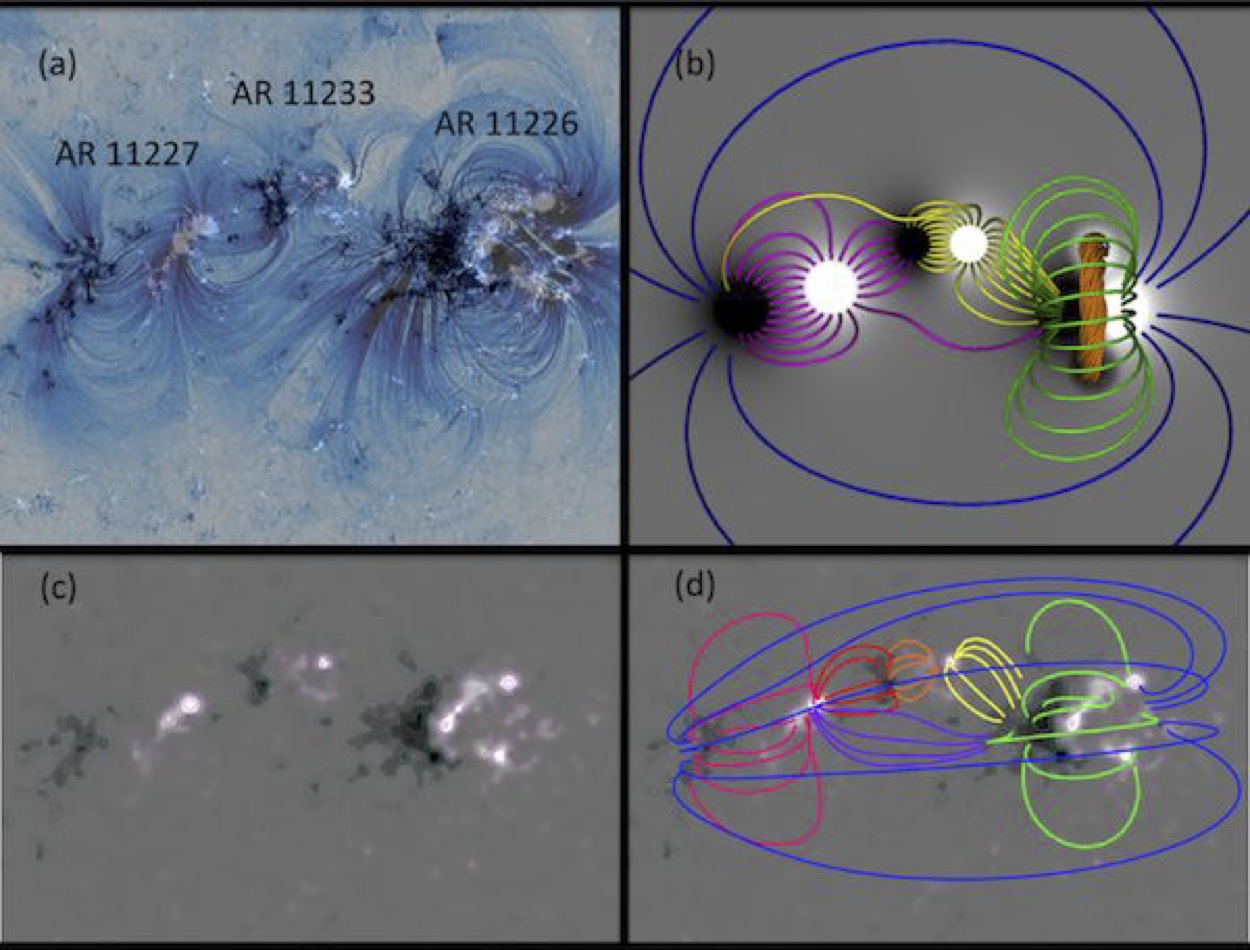}
\caption
{Observed and modelled/simulated magnetic configuration of the three-AR complex on 2011 June 3, four days before the CME eruption from AR 11226, giving a top view of the AR complex. (a) {\sl SDO}/AIA 171 \AA~ reverse colour scale image over a co-aligned {\sl SDO}/HMI magnetic field map of the three neighbouring active regions. (b) Corresponding magnetic configuration in the simulation. Black/white indicates negative/positive magnetic polarity. The pre-eruption magnetic flux rope in the rightmost/westernmost active region is shown in gold. (c) Smoothed {\sl SDO}/HMI magnetogram. (d) Potential magnetic extrapolation with each colour representing a different quasi-connectivity domain.
}
\label{fig_config}
\end{figure}

The 2011 June 7 eruption produced a fast CME ($\approx$ 1250 km\,s$^{-1}$), a large-scale EUV wave, an M2.5 flare, and fireworks comprised of dense blobs of filament material that fell back towards the Sun, impacting over almost a quarter of the solar disk (see the multi-wavelength {\sl SDO}/AIA movie [movie1.mov] in the online version).  Various aspects of this eruption have been analyzed so far, including the resulting global wave, the relationship between hard X-rays and UV radiation from the flare arcade, the density of back-falling plasma blobs and their formation by the Rayleigh-Taylor instability, the kinetic energy of the blobs and the heating caused by their impact -- the latter was likened to processes taking place during stellar accretion \citep{Li12,Innes12,Cheng12,Williams13,Inglis13,Carlyle14,Gilbert13,Reale13}. Furthermore, there has been a detection of enhanced gamma-ray ($>$ 100 MeV) activity from the Sun by the Large Area Telescope on the Fermi Gamma-Ray Space Telescope in the period of 06--12~UT, i.e., during the eruption \citep{Tanaka11}, as well as of solar neutrons with the International Space Station's FIBer instrument (part of the {\sl Space Environment Data Acquisition equipment (SEDA)}; \citealt{Muraki12}).

\begin{figure}[t]
\centering
\includegraphics[width=1.\linewidth]{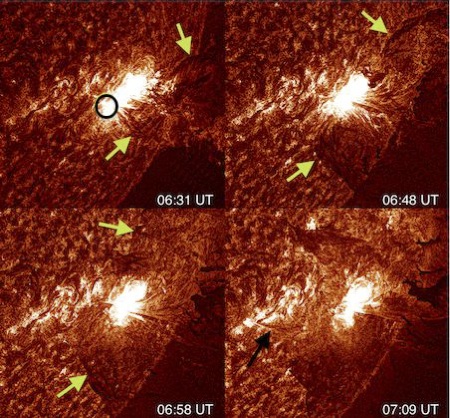}
\caption
{
{\sl SDO}/AIA 304 \AA~ images show the evolution of the massive filament eruption in the lower corona on 2011 June 7 from AR 11226. Note the large lateral expansion of the erupting filament marked by light green arrows and back-falling plasma blobs seen in absorption above the limb, which made this CME so spectacular.  Along the South-East (SE) flank of the erupting filament, dark filament material is flowing back towards its SE footpoint in AR 11226 (black circle). Note a new connection at 07:09 UT, which channels filament material towards the negative polarity footpoint of AR 11227 (black arrow). The formation of this new connection is better seen in the AIA multi-wavelength movie in the electronic version (movie1.mov).
}
\label{fig_304}
\end{figure}

\begin{figure}[htpb]
\centering
\includegraphics[width=0.9\linewidth]{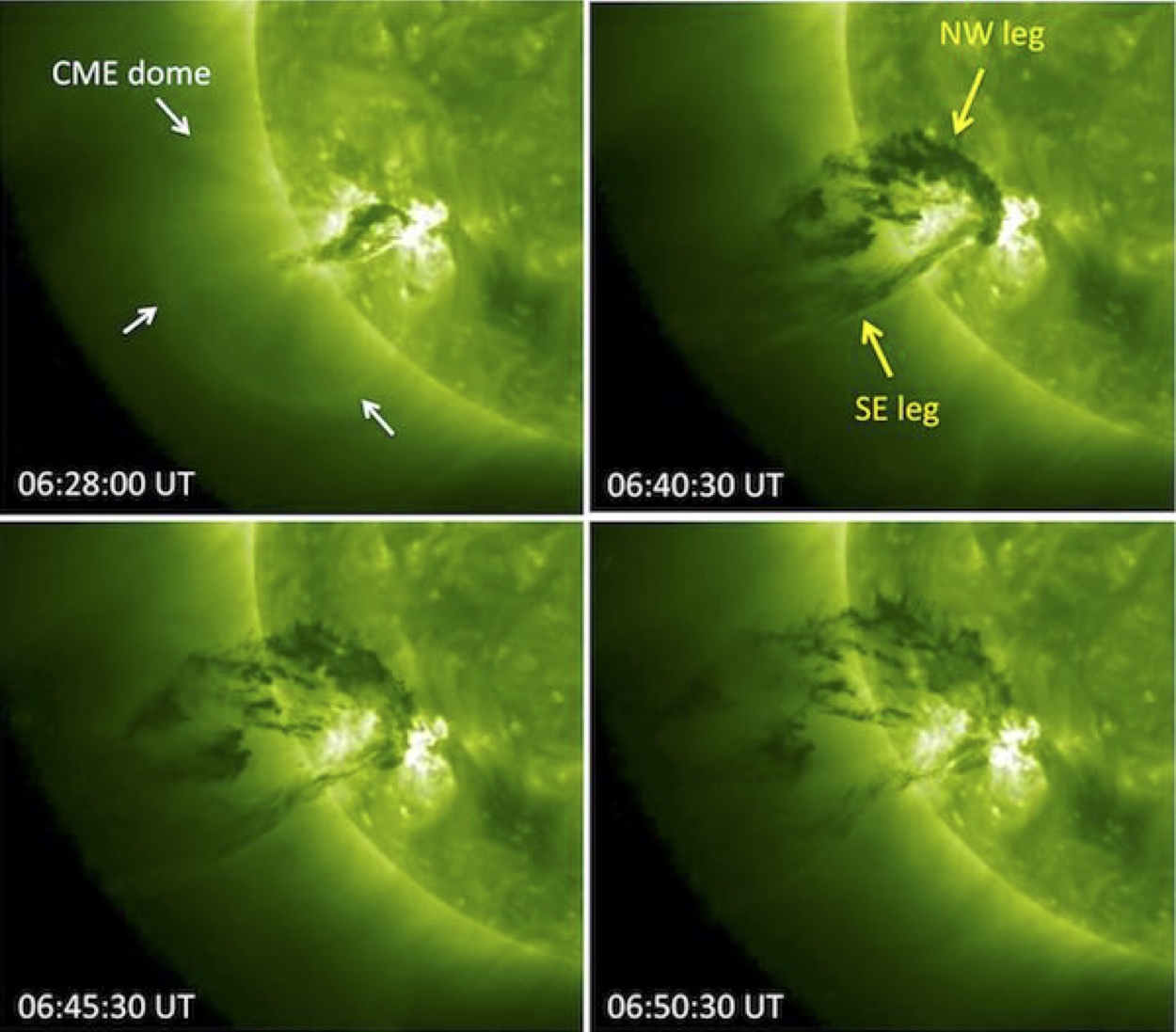}
\caption
{
STEREO-A SECCHI/EUVI 195 \AA~ images of the filament eruption/CME on 2011 June 7, when STEREO-A was 94.9$^{\circ}$ ahead of Earth allowing us to see the event from a western-side perspective. The series of images show the bright dome of the forming CME (white arrows) and the evolution of filament eruption in its core with yellow arrows marking its legs.
}
\label{fig_STa}
\end{figure}

The CME originated in a complex of three adjacent active regions (ARs 11226, 11227, and 11233, see Figure \ref{fig_config}(a)) in the south-western quadrant of the Sun (as seen from Earth, c.f. Figure \ref{fig_304}) and carried an unusually massive erupting filament in its core (Figure \ref{fig_STa}). As the filament was rising in the solar atmosphere from AR 11226, it showed fast lateral expansion, with its flanks reaching as far as the neighboring ARs' magnetic loops. In order to better understand the topology of the eruption, we have carried out a data-constrained MHD simulation, which demonstrates that the lateral expansion of the CME flux rope led to magnetic flux pile-up, yielding the dynamic formation/intensification of current sheets where magnetic reconnection between the flux rope and the magnetic field of the neighboring ARs took place. 

For the first time, {\sl SDO}/AIA observations revealed clear and unmistakable signatures of magnetic reconnection between an expanding CME and a neighboring AR, including the re-direction of back-flowing filament material towards remote locations in one of the neighbouring ARs (AR 11227). Furthermore, they showed brightening of coronal plasma in the environment of the reconnection region due to the presence of unusually high densities there. These observations provide evidence that during the expansion of an erupting flux rope/CME, nearly instantaneous magnetic reconnection can occur with ambient magnetic field, illustrating how the solar atmosphere is reconfigured during large-scale eruptions. 

The article is organized as follows. In Section 2 we describe the observations used, the  modelling methods, and the MHD simulation setup. In Section 3 we present the observational, modelling, and simulation results. Finally, in Section 4 we discuss the results and present our conclusions.

\section{Observations, QSL computations, and MHD simulation setup}

\subsection{Observations}
\label{obs}
For this work we use {\sl Solar Dynamics Observatory} \citep[{\sl SDO}:][]{Pesnell12} Atmospheric Image Assembly \citep[AIA:][]{Lemen12}, which takes images of the full solar disk with a resolution of about 0.6$\arcsec$ pixel$^{-1}$ through 10 filters, selected to include specific strong lines in the corona and continuum emission from the lower chromosphere. We investigate cold absorbing and hot emitting structures. Although we checked images in all 10 channels, for this analysis data from the 304, 171, 193, and 211 \AA~ channels were the most suitable. The cadence was on average 12 s in the 171, 193, and 211 \AA~ channels, and 60 s in the 304 \AA~ channel. The 304, 171, 193, and 211 \AA~ passbands are dominated by ions formed at temperatures around 5 $\times$ 10$^{4}$, 6.3 $\times$ 10$^{5}$, a combination of 1.2 $\times$ 10$^{6}$ \& 2 $\times$ 10$^{7}$ K, and 2 $\times$10$^{6}$ K, respectively.

Additionally, 195 \AA~ EUVI images are used from SECCHI \citep{Howard08} on STEREO-A, which was 94.9$^{\circ}$ ahead of Earth, and observed the eruption in the SE quadrant close to the limb. 

We also use line-of-sight (LOS) magnetic field data from the Helioseismic and Magnetic Imager \citep[HMI:][]{Schou12,Scherrer12} onboard {\sl SDO}. The LOS magnetograms are taken in the 6173.3 $\pm$ 0.1 \AA~ Fe {\sc I} line, with an angular resolution 0.5$\arcsec$ pixel$^{-1}$, a cadence of 45~s, and a noise level of 10 G.

\subsection{Magnetic extrapolation and magnetic topology computations}
\label{qsl_method}
\begin{figure}[t]
\centering
\includegraphics[width=1.\linewidth]{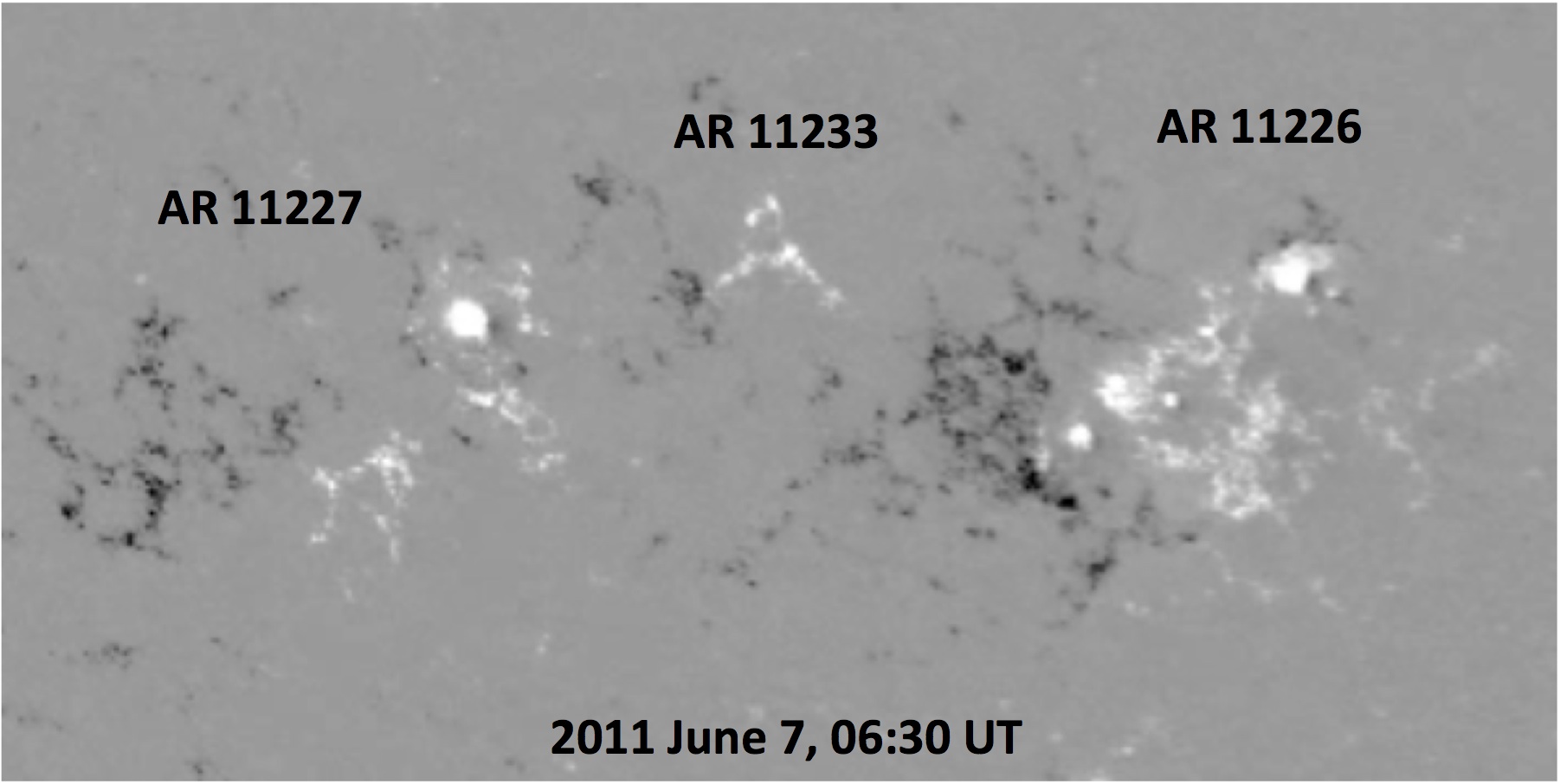}
\caption
{\textsl{SDO}/HMI observation of the magnetic fields (white/black represents positive/negative polarity) of the three-AR complex on 2011 June 7 at 06:30 UT, at the peak of the M2.5 flare, which accompanied the CME analysed here. Comparing the photospheric flux distribution to that four days prior to the eruption shown in Figure \ref{fig_config}(c), it is clear that in spite of the unusually high amount of flux cancellation in AR 11226, the overall photospheric flux distribution did not significantly change.  For the full magnetic evolution leading up to the eruption, see the \textsl{SDO}/HMI movie [movie2.mov]) in the online version.
}
\label{fig_mag}
\end{figure}

As the CME eruption occurred close to the western limb (S22, W53), we used {\sl SDO}/HMI data four days prior to the CME on 2011 June 3 at 20 UT for the analysis of the magnetic connectivities and topology in the three-AR complex including the CME source region AR 11226, and two other ARs (ARs 11227 and 11233) on its East side (Figure \ref{fig_config}(a)). During the four days leading up to the eruption a significant amount of magnetic cancellation took place within AR~11226, accompanied by weaker flux cancellation in the other two ARs (see Figure \ref{fig_mag} and the \textsl{SDO}/HMI movie in the online version [movie2.mov]). Since all three ARs were in their decaying phase, their overall photospheric flux distribution did not change substantially, and their relative positions remained fairly constant, it is reasonable to assume that the overall topology of the magnetic connections between the ARs did not significantly change in these four days.
 We carried out a potential-field extrapolation (Figure \ref{fig_config}(d)) of the {\sl SDO}/HMI LOS magnetogram (Figure \ref{fig_config}(c)) applying the method of \cite{Schmidt64}. 
 
We also computed the location of quasi separatrix layers \citep[QSLs:][]{Demoulin96} in the three-AR complex on 2011 June 3. The location of the QSLs is determined by computing the values of the squashing degree, Q,  \citep{Titov02,Titov07} in the domain using the method described in \citet{Pariat12}. The horizontal plane one pixel above the boundary (at $z \sim 4.5$ Mm) is taken as the reference plane. This enables us to partly remove the QSLs related to  small magnetic polarities and localised bald patches \citep[as  in][]{Savcheva12}. We compute and plot 
the QSLs on this plane using an initial grid with a mesh size of 0.7$\arcsec$ ($\sim 0.5$ Mm) with the Q values eventually computed on an adapted mesh of length $5.5 \times 10^{-3}$$\arcsec$.  Regions where one of the footpoints of the field  lines has a magnetic field intensity smaller than $10$ G are also not considered. As a result, the large-scale 
QSLs separating the different connectivity domains are better highlighted. We also performed computation along a vertical cut of the domain, in an $xz$ plane at $y=-430\arcsec$. The initial grid for that vertical computation has a mesh size of $\sim 1.5\arcsec$ ($\sim 1$ Mm) and reaches $1.1 \times 10^{-2}$$\arcsec$ at the maximum resolution of the Q computation.

\subsection{MHD simulation}
\label{sim_method}
To further investigate the interaction between the CME and the neighboring AR 11227, we perform an idealized, three-dimensional MHD simulation. Although the parameters of the initial configuration are guided by the observed values, we do not attempt to match all numbers closely, as the simulation is intended to serve as a proof of concept, not to reproduce the observed eruption as closely as possible. We employ the coronal flux rope model of \cite{Titov99}, hereafter TD, to construct the initial magnetic field in the simulation. The main ingredient of the TD model is a toroidal current channel of major radius $R$ and minor radius $a$, positioned such that the symmetry axis of the torus is located at a depth $d$ below the photospheric plane. The outward directed Lorentz self-force (or ``hoop force'') of the current ring is balanced by the force provided by the external field. This field is created by a pair of sub-photospheric point sources, $\pm q$, which are placed at the symmetry axis, at distances $\pm L$ from the torus center (see Figure~2 of TD). The resulting force-free coronal equilibrium consists of an arched and line-tied flux rope embedded in bipolar potential field.

We build an initial magnetic field that consists of three bipolar ARs, of which the western one (representing the CME source AR 11226) contains a TD flux rope, while the other two ARs contain purely potential fields. The chosen parameter values ensure that the rope is initially stable with respect to the helical kink \citep{Torok04} and torus instabilities \citep{Kliem06}. The AR centers are placed such that they correspond to the relative positions of the observed ARs. The distance between the two sub-photospheric point sources for each AR is also guided by the observations, and the signs of the point sources are chosen according to the observed polarities. In order to reflect that AR 11226 exhibited the largest flux, the point sources of the western AR are 25~percent stronger than those in the other ARs. 
Figure \ref{fig_config}(b) shows the resulting magnetic configuration which is similar to the potential field extrapolation of the three-AR complex in Figure \ref{fig_config}(d).

We integrate the zero-$\beta$ compressible ideal MHD equations, neglecting thermal pressure and gravity, which is justified in the low corona where the Lorentz force dominates. The MHD variables are normalized by the initial TD torus axis apex height, $R-d$, the maximum initial magnetic field strength $\bf{B}_{\rm 0,max }$, the maximum Alfv\'en velocity, $v_{a\rm 0,max }$, and derived quantities. The Alfv\'en time is $\tau_{a} =  (R - d )/v_{a\rm 0,max}$. We use a nonuniform Cartesian grid of size $[-25, 25] \times [-25, 25] \times [0, 50]$ with a resolution of $\simeq0.04$ in the flux rope area. The initial density distribution is $\rho_{0}(\bf{x}) = |\bf{B}_{0}(\bf{x})|^{3/2}$, such that $v_{a}$(x) decreases slowly with distance from the flux concentrations. For further numerical details we refer to \cite{Torok03}.

We first relax the system for 58\,$\tau_{a}$  and reset the time to zero. Then the eruption of the flux rope is triggered by imposing localized, strongly sub-Alfv\'enic converging flows at the bottom plane, similar to the flows applied in \cite{Torok11a}. The  flows slowly drive the polarities surrounding the flux rope toward the local inversion line, yielding a quasi-static expansion of the flux rope's ambient field. As a result, the flux rope slowly rises until it reaches the critical height for the onset of the torus instability, which then produces an eruption \citep[see also][]{Torok11b}. 
Such converging motions, which were indeed observed to be persistent and strong, provide an efficient way to destabilize the flux rope and produce an eruption.  As the SDO/HMI movie [movie2.mov] shows, the typical converging motions of opposite polarities induced by flux dispersion were enhanced by the fragmentation of the leading positive spot during the seven days prior to the eruption and the highly unusual motion towards the main magnetic inversion line of one of the large spot fragments, which gradually cancelled there, removing equal amount of negative polarity.

Although we solve the ideal MHD equations, numerical diffusion occurs in regions of high current densities, enabling magnetic reconnection of field lines. As the diffusion in the simulation is much higher than expected in the solar corona, we refrain from a quantitative investigation of the reconnection scaling properties. In spite of this limitation, the simulation produces solutions which show an excellent match with the observed event (see Sect.\,\ref{ss:MHD_results}).

\section{Results}
\label{results}

\subsection{General description of the event}
The unusually massive filament considered in this paper formed in AR~11226, which was part of a complex of three adjacent decaying bipolar regions. It erupted at about 06:15 UT on 2011 June 7. The accompanying M2.5 flare had a start time of 06:16 UT and peaked at 06:30 UT. A dome-like bright front (Figure \ref{fig_STa}) formed by about 06:26 UT, which expanded laterally reaching its maximum speed of $\sim$450 km s$^{-1}$ by 6:36 UT after which the radial expansion started slowing down \citep[for details see][]{Cheng12}. The erupting filament in the core of this expanding dome-like shell, well behind the bright front, also expanded in the lateral direction while erupting (Figures \ref{fig_304} and \ref{fig_STa}). This lateral expansion brought it in contact with neighbouring magnetic structures, in particular, with magnetic field lines/loops of AR 11227 (Figure \ref{fig_config}). While the erupting flux rope was lifting and accelerating the bulk of the filament, many plasma blobs formed via the Rayleigh-Taylor (RT) instability \citep{Innes12,Carlyle14} and began to detach from the filament. This occurred in the particularly massive parts of the filament, i.e., in its top part and in the North-West (NW) leg. These blobs obviously did not follow the fieldlines threading the filament in their return to the solar surface. However, their descent appeared to be influenced/guided by magnetic field, implying that they were dragging field lines with them. On the other hand, along the straighter South-East (SE) leg of the filament cool material was mainly flowing back along the fieldlines towards the SE footpoint (Figure \ref{fig_STa}), developing only minor indications of the RT instability. An interaction of the SE flank with external magnetic structures, via magnetic reconnection, became apparent in AIA images when the formation of a new branch of downward flowing filament material was observed (see a snapshot of the well-developed process at 07:09~UT in Figure \ref{fig_304} and full detail in Figures\,\ref{fig_density}--\ref{fig_stackplot2} as well as the online multi-wavelength movie1.mov).
Dense, cool filament plasma originally contained within the erupting flux rope was re-directed towards remote magnetic footpoints in the neighbouring AR~11227, thus unambiguously tracing the change of large-scale magnetic connectivity.

\subsection{Density of the redirected back-flowing filament material}
\label{density}
\begin{figure}[t]
\centering
\includegraphics[width=0.75\linewidth]{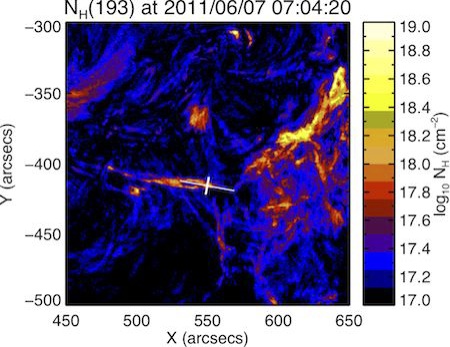}
\includegraphics[width=0.75\linewidth]{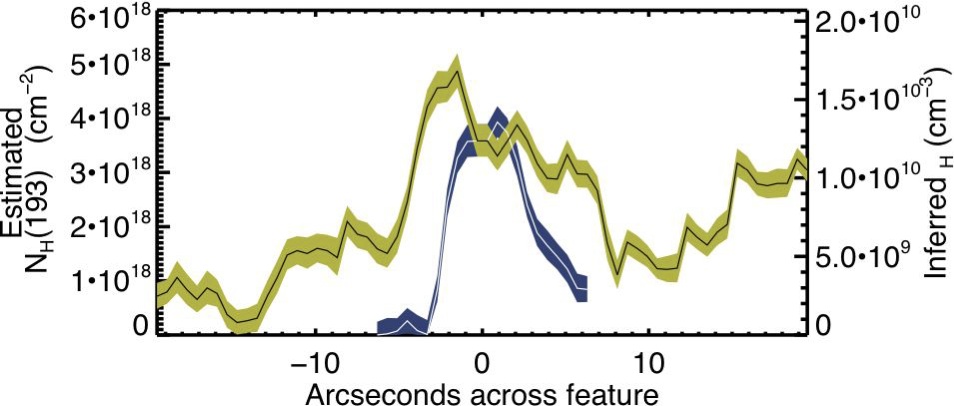}
\caption
{(a) Column electron density map (cm$^{-2}$) and (b) column- and inferred volumetric electron density (cm$^{-3}$) in the vicinity of the magnetic reconnection region (zoom on the ``new connection'' in Figure \ref{fig_304}) along the slices shown by the cross. The black/white line with 1$\sigma$ green/blue envelope corresponds to the along/across the feature profile position in the top panel. For the volumetric density calculation the depth was assumed to be equal to the width ($\sim$4$\arcsec$). The density fluctuations seen along the longer profile are indicative of the presence of small blobs in the filament plasma.
}
\label{fig_density}
\end{figure}

The back-flowing material from the ejected filament is optically thick, in that it obscures emission from the corona behind it through photoionization ({\it i.e.}, photoelectric) opacity. The strength of this absorption allows the column density of the absorbing species -- {\hi}, {\hei} and {\heii} -- to be determined. In principle the optical depth of this absorption may be measured by comparing the intensity of pixels containing this prominence matter with pixels either side; this is known as a spatial-interpolative approach and has been exploited to great effect by \cite{Kucera98}, \cite{Golub99} and \cite{Gilbert05,Gilbert06}, and is appropriate in cases where the background emission has to be modelled because it is not seen directly.  Here, however, we take advantage of the fact that the low-ionization material sweeps through the AIA image, and use images just before its arrival to model the background radiation field: this constitutes a temporal-interpolative approach. In doing so, we can more accurately estimate the column depth of material contained in the back-flowing material just before it enters the region of interest. Under the assumptions that the foreground EUV emission is negligible, and that the depth of the plasma blob/thread is equal to its full-width at half-maximum in the plane of the sky (approximately 4$\arcsec$), a lower limit of the electron density in the redirected filament plasma in the vicinity of the reconnection region was estimated to be $10^{10}$ cm$^{-3}$ \citep[Figure \ref{fig_density}; for details see][]{Williams13}. This is more than one order of magnitude larger than the typical coronal density of bright loops.  

\subsection{Brightening around the reconnection region}

\begin{figure*}[t]
\centering
\includegraphics[width=0.84\linewidth]{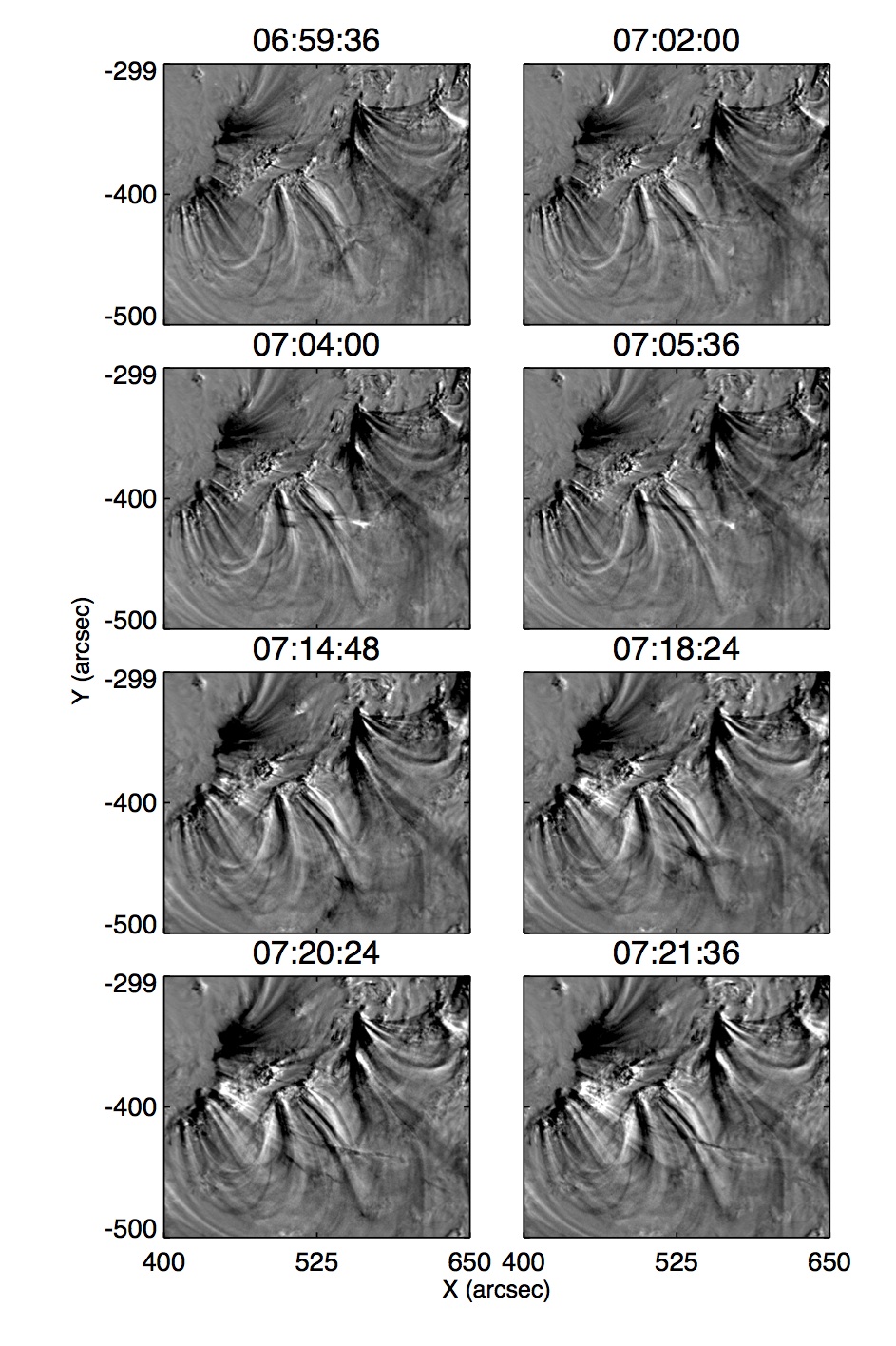}
\caption
{
AIA 171 \AA~ zoomed base difference images of the reconnection region between ARs 11226 and 11227. The base image was taken at 06:45:12 UT. The SE flank of the erupting filament enters the FOV from the West at about 06:48 UT, then by about 07:00 UT a new eastward branch forms. Sporadic heated patches are present from the beginning of the restructuring in the region. The brightest emission around the reconnection region starts at 07:02:24 UT, reaches a peak at 07:04:24 UT, and fades by 07:07:36 UT, while shifting westward. A second ``train'' of dark material arrives in the FOV from the SW at 07:11 UT. The blobs transform into long, thin threads by about 07:22 UT as they are falling towards their footpoints in AR 11227. \\ 
}
\label{fig_brightening}
\end{figure*}

The re-direction of back-flowing filament plasma started at around 06:57 UT, and by 07:00 UT plasma was flowing towards the negative-polarity footpoints of AR~11227 in well-developed threads (see the AIA 171~\AA\ data in Figure \ref{fig_brightening} and the accompanying movie1.mov). Around the location of the sharp change in plasma flow direction, some sporadic brightenings have appeared by 07:00 UT, which intensified with time. The brightest emission was observed in the 171 \AA~ channel between 07:04 and 07:06 UT.  The appearance of emission in EUV around the redirection/reconnection site clearly indicates that in-situ heating was occurring there.  Particles accelerated during the reconnection process were in an unusually dense coronal environment, and transferred their kinetic energy through Coulomb collisions to the filament plasma. Heating by slow-mode shocks (i.e. Petschek shocks) is also expected to take place. 

\begin{figure}[t]
\centering
\includegraphics[width=0.95\linewidth]{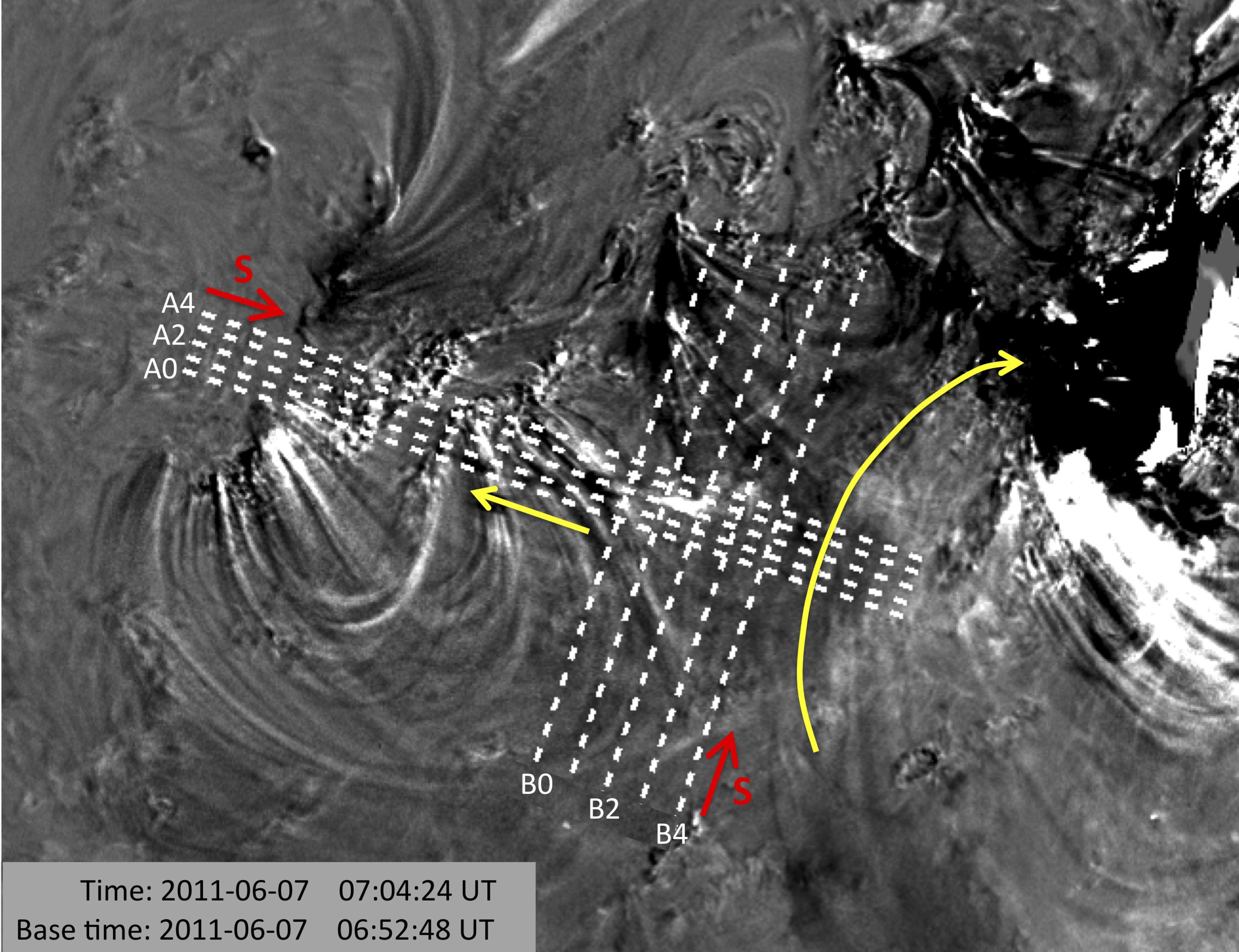}
\includegraphics[width=0.95\linewidth]{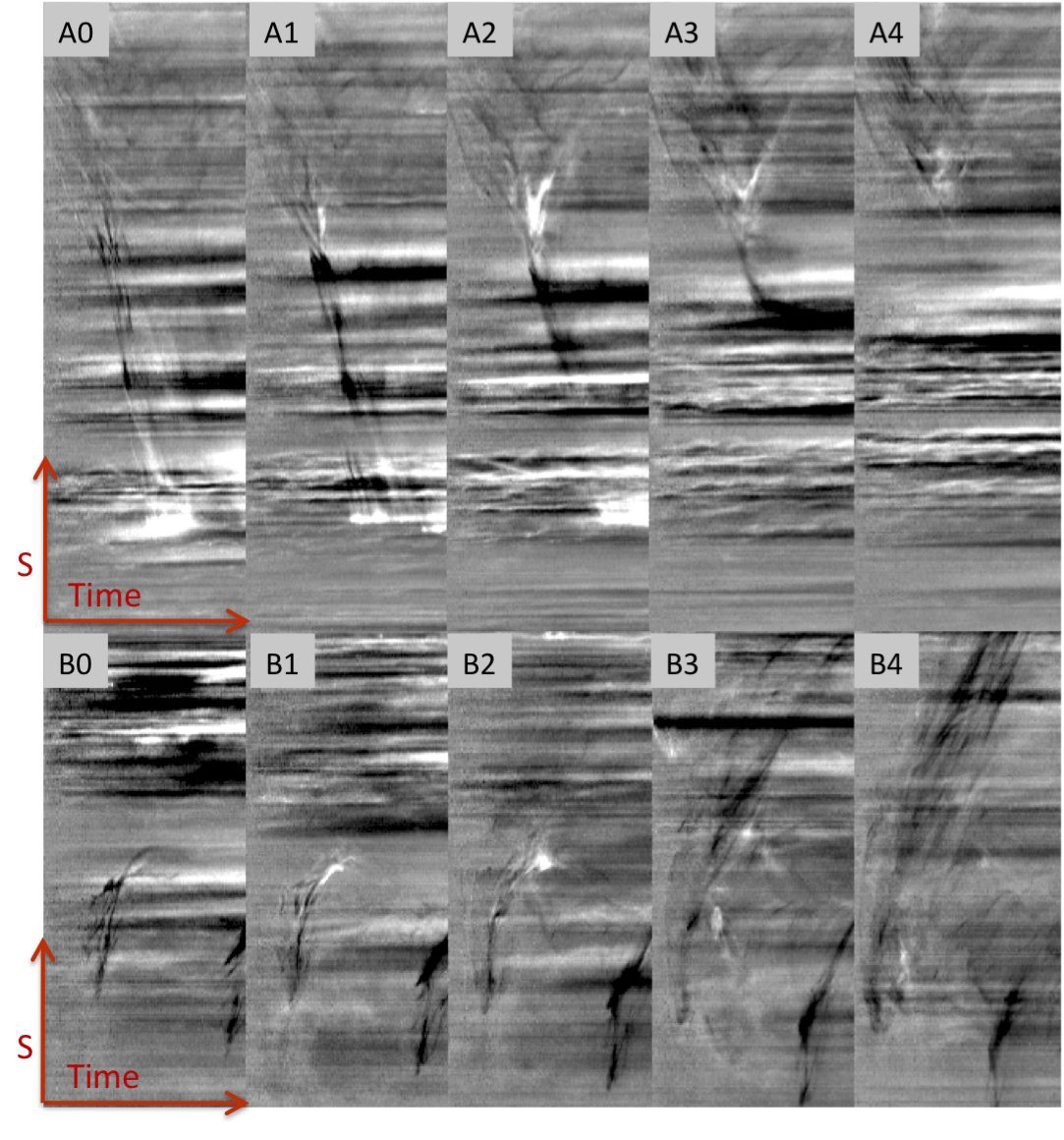}
\caption
{
Series of stack-plots in the AIA 171 \AA~ band base-difference images. The top panel identifies the cut directions and locations. The long curved yellow arrow indicates flow direction within the SE leg of the erupting filament, and the short straight yellow arrow gives the direction of the redirected plasma. In the middle (A-cuts) and bottom (B-cuts) panels the horizontal axis is time from 06:53 to 07:19 UT, the vertical dimensions (marked with a red arrow and S) are 177 Mm (240$\arcsec$) for the A cuts and 132 Mm (180$\arcsec$) for the B cuts. The A and B cuts are approximately along, and perpendicular to, the different threads of the redirected down-flowing plasma. 
The stack-plots A0, A1 clearly indicate that cold, absorbing plasma is flowing towards the magnetic footpoints in AR 11227 and becomes heated when it reaches lower height, higher-density layers. They show little signature of chromospheric evaporation (no V-shaped bright features around the impact points of the flow).   
}
\label{fig_stackplot1}
\end{figure}

\begin{figure}[t]
\centering
\includegraphics[width=0.9\linewidth]{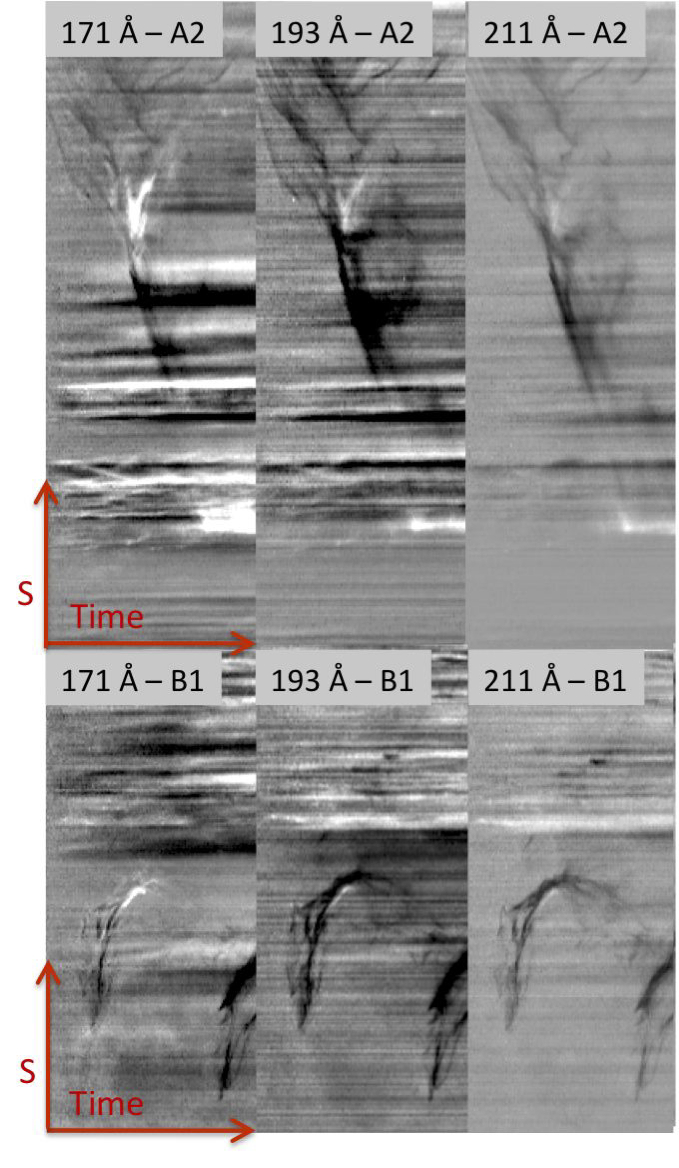}
\caption
{
Stack-plots along the A2-B1 lines shown in Figure \ref{fig_stackplot1} (upper panel) in three AIA channels: 171 \AA~ (6.3 $\times$ 10$^{5}$ K), 193 \AA~ (1.2 $\times$ 10$^{6}$ \& 2 $\times$ 10$^{7}$ K) and 211 \AA~ (2 $\times$ 10$^{6}$ K). Time goes from left to right (06:53--07:19 UT in all three channels). The vertical dimensions (marked with a red arrow and S) are 177 Mm (240$\arcsec$) for the A cuts and 132 Mm (180$\arcsec$) for the B cuts.The plots show that heating around the reconnection region as well as around the magnetic footpoints of the reconnected loops are the strongest in the 171 \AA~ channel, and it is also present at million-degree coronal temperatures. The dimensions of the cuts are the same as in Figure \ref{fig_stackplot1}.
}
\label{fig_stackplot2}
\end{figure}

Stack-plots of two series of narrow cuts approximately along (A-cuts), and perpendicular to (B-cuts) the threads of the down-flowing plasma redirected by reconnection is shown in Figure \ref{fig_stackplot1}, revealing the evolution of plasma behavior  around the reconnection region. In the top panel in Figure \ref{fig_stackplot1} the location of the cuts are indicated over an AIA 171 \AA~ base-difference image. The middle panels show stack-plots along the series of A-cuts, the bottom panel along the B-cuts. 

Within the SE leg of the erupting and laterally expanding filament, plasma blobs were swirling in clockwise direction towards the SE (negative) magnetic footpoint in AR 11226. As the motions were quite complex and formed part of a complex evolution with blobs/threads of different heights overlapping in the images, the interpretation of the stack-plots is not straight-forward. 

The B-cuts run roughly parallel to the original (projected) flow direction along the SE leg of the erupting filament (indicated by a long curved arrow in Figure \ref{fig_stackplot1} top panel), and the B3--B4 cuts capture the proper motion of some of the blobs. However, what we see in the B0--B2 cuts is not real plasma motion, as different blobs, presumably on different field lines,  enter and leave the cut directions in succession, forming an apparent continuous thread. Eventually, some of the blobs enter the reconnection region and get heated. As the reconnection region is shifting along the B-cut direction, the stack-plots capture that motion as a bright track.  

The cuts A0--A1 capture the entire path of the redirected filament plasma, which is moving along reconnected field lines. These cuts show many thin threads, which were formed at different times and give the impression of torsional motion around one another (A1). 
The cold plasma flowing towards the footpoints became heated in lower height, higher-density layers. As there are no bright V-shape features around the footpoints in the stack-plots of the cuts A0--A1, chromospheric evaporation was apparently not strong. 
 
In Figure \ref{fig_brightening}  the bright reconnection region (i) extends in length along the A-cut direction and (ii) between 07:04:24 and 07:06:48 UT presents a significant proper motion along the A-cut direction. These both are  manifested in the A-cut stack-plots. 
In the upper (filament-side) part of the A-cut stack-plots, seen above the bright reconnection region in Figures \ref{fig_stackplot1} and \ref{fig_stackplot2}, the most striking features are V-shape threads. The downward branch of the V shows expanding (southeastward) motion of the filament leg before reconnection. The upward branch of the V shows northwestward-directed
downflow along the non-reconnected part of the flux in the filament leg accelerating due to gravity into AR 11226.


A similar V-pattern appears in the stack-plots presented by \cite{Su13} for the 2011 August 17 limb flare, interpreted as reconnection inflow and outflow in the flare using coronal loops as tracers. However, in our case the picture was made more complex by the presence of cold plasma blobs as tracers. These plasma blobs populated only segments of a field line and even \textit{moved} along the magnetic field lines. Therefore, the stack-plots of the reconnection region presented by \cite{Su13} and our Figures \ref{fig_stackplot1} and \ref{fig_stackplot2} are not directly comparable.

\begin{SCfigure*}
\centering
\includegraphics[width=1.5\linewidth]{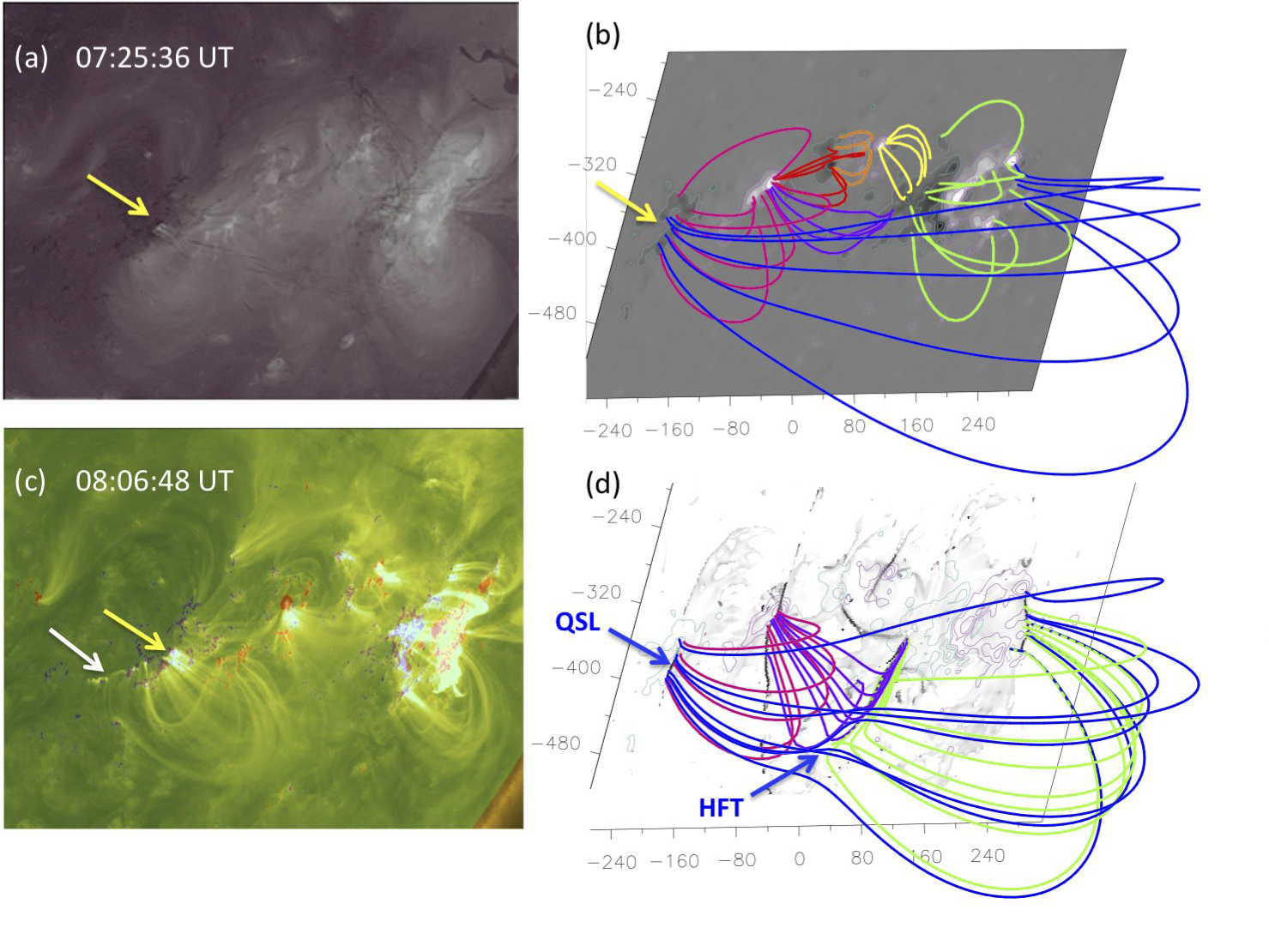}
\vspace{-0.5cm}
\caption{Footpoint locations of the reconnected fieldlines and quasi-separatrix layer (QSL) locations. (a) AIA 211 \AA~ image showing the first set of bright footpoints (yellow arrow), and (c) AIA 171 \AA~ image showing extension of the bright ribbon towards the weaker fields (white arrow). (b) The potential field extrapolation of the 2011 June 3 (c.f. Figure \ref{fig_config}(d)) as well as  (d) the QSL map computed from it, rotated to the time of the event. A comparison of these panels indicates that the footpoints of the reconnected loops are located around a QSL separating drastically different magnetic connectivities: field lines connecting to the leading positive polarity of AR 11227 (in red) and large-scale field lines connecting to the leading positive polarity of AR 11226 (in blue). The computed magnetic topology also reveals the presence of a hyperbolic flux tube (HFT) between the magnetic connectivity domains of AR 11226 and AR 11227. }
\label{fig_qsl}
\end{SCfigure*}

The multi-wavelength capability of {\sl SDO} allows a rough diagnostic of temperature dependence of plasma heating in the vicinity of the reconnection region and the footpoints. The most prominent brightening is seen in the 171 \AA~ waveband, indicating that the bulk of the plasma has been heated to $\approx 6 \times 10^{5}$ K. This is the case both for the reconnection region and the footpoints (see Figure \ref{fig_stackplot2}). The emission of the reconnection and footpoint regions is weaker in the 193 \AA~ waveband, which images plasma at $\approx 1.2 \times 10^{6}$ K. The weakest emission is seen in the 211 \AA~ waveband which images plasma at $\approx 2 \times 10^{6}$ K. This indicates that these regions are only heated to the lower end of the coronal temperature range. 

The emission region seen in Figure \ref{fig_brightening} is shifting towards the West, which is also seen in Figure  \ref{fig_stackplot1} (middle panels, A2--A3) as an upward propagating branch of the emission region. 
This westward shift of the bright emission region might result from successive reconnections progressing toward the core of the filament flux rope, but the relatively high displacement velocity of $\sim90$~km~s$^{-1}$ makes it more likely that the reconnection region shifts westward. Such a shift could be the result of the large-scale flows from the encounter of the filament flux with the flux of AR 11227. These flows have a downward component inside the leg of the filament, which drives the interaction. This represents a flow along the current sheet and may cause a shift of the reconnection X-line in the direction of the flow, westward in projection. Such a flow is not normally included in reconnection models.

\subsection{Magnetic topology and footpoint locations}
How can we put further constraints on the identity of the magnetic loops which reconnect with the filament's expanding flux rope? We use the tools of magnetic topology analysis: a magnetic extrapolation and a map of quasi-separatrix layer locations for this.  

As the location of footpoint brightenings (Figure \ref{fig_qsl}(a,c)) provides insight into the reconnecting loops' identity, we use a magnetic extrapolation (Figure \ref{fig_qsl}(b)) and a map of QSLs (Figure \ref{fig_qsl}(d)) in the three-AR complex to explore this question. QSLs are locations where there is a large gradient (i.e. drastic change) in magnetic connectivity (for the method, see Section \ref{qsl_method}). Comparing the four panels in Figure \ref{fig_qsl} it is apparent that the impact-brightened footpoints  are over the negative polarity of AR 11227. This is where two sets of loops originate: one set of medium-long loops (in red) connecting to the leading positive polarity of AR 11227, and another set of long loops (in blue) connecting to the leading positive polarity of AR 11226 (Figure \ref{fig_qsl}(c)). Between these two sets of loops, a QSL is present (Figure \ref{fig_qsl}(d)). The impact-brightened footpoints are on both sides of this QSL, implying that magnetic field lines of the erupting filament reconnected with both sets of these magnetic loops.

\begin{SCfigure*}
\centering
\includegraphics[width=1.5\linewidth]{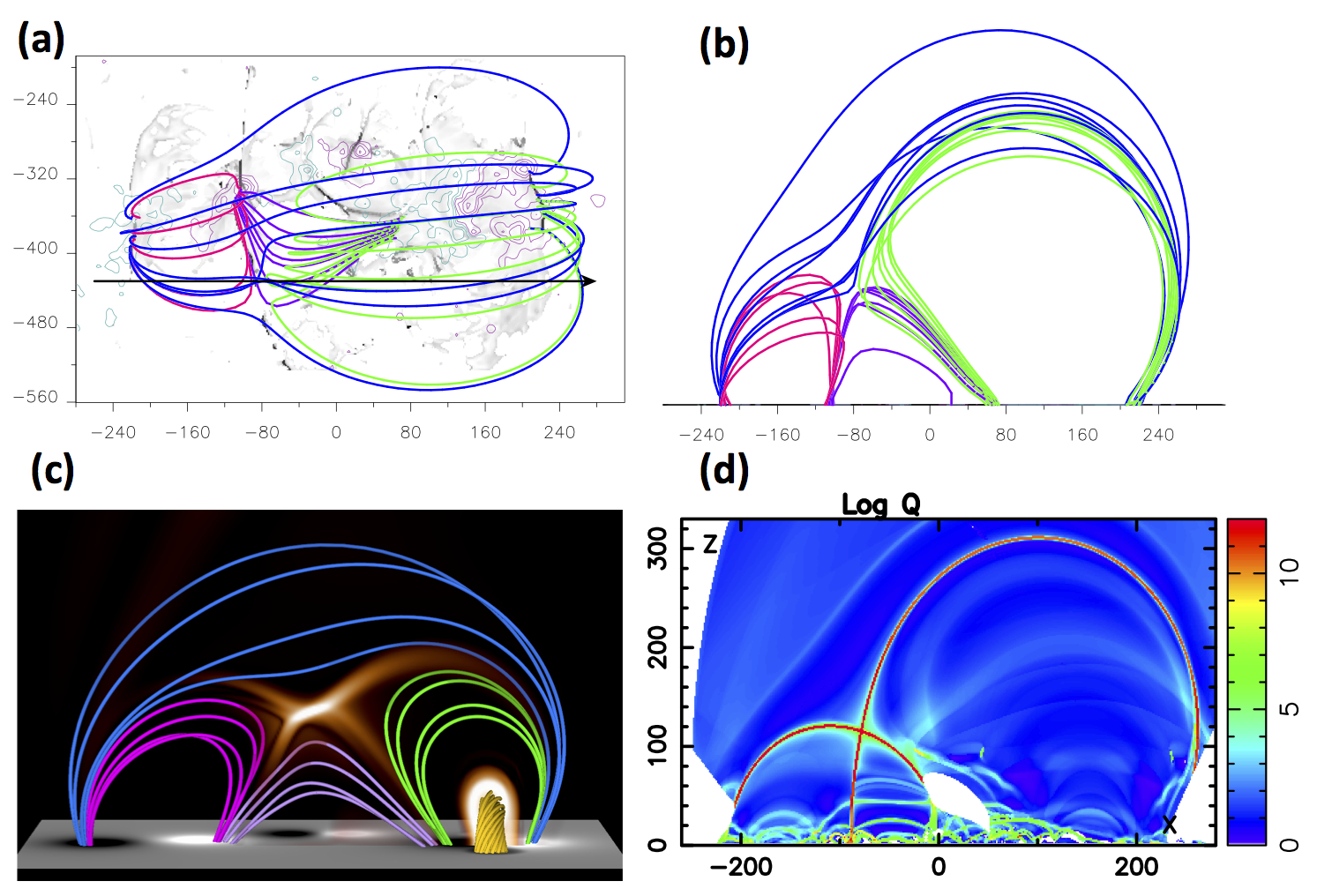}
\caption{Quasi-separatrix layer (QSL) locations including a hyperbolic flux tube (HFT) in the three-AR complex computed from the {\sl SDO}/HMI data (a, b, d), and electric current concentrations in the MHD simulation (c). The computed magnetic topology is shown in (a) top and (b) side views. Different magnetic connectivity domains are marked with different colours. The black arrow in panel (a) (at Y$\approx$-430$\arcsec$) indicates the location of the vertical cut shown in panel (d). The scale indicated in the panels is in arcsec. Current density enhancements in the MHD simulation at $t=0$ after the
relaxation (c) correspond to the QSLs in the extrapolation (d). Both (c) and (d) show a cross section of an HFT (X-type structure) which extends predominantly perpendicularly to the
plane of the figure (compare with Figure \ref{fig_qsl}(d)). 
}
\label{fig_qslcut}
\end{SCfigure*}

We found a group of null points over a quiet-Sun area outside of the complex, at a height of 4 Mm at around y=-480$\arcsec$ along the QSL extending southward from the leading polarity of AR 11227 (Figure \ref{fig_qsl}(d)). These quiet-Sun nulls did not seem to play any significant role in the reconnection process described here, although at the start of the reconnection event a narrow channel of southward-flowing filament material formed in the direction of these nulls.  

The topology analysis revealed the existence of a hyperbolic flux tube \citep[HFT:][]{Titov02} at a height of about $\approx$80 Mm (110$\arcsec$), as shown in Figure \ref{fig_qslcut}(d), between the connectivity domains of ARs 11226 and 11227 (c.f. Figures \ref{fig_qsl} and \ref{fig_qslcut}), at the location where the reconnection is observed to take place. It is most likely that the expanding flux rope of the CME/filament provided a forcing along this HFT, which led to the observed reconnection. 

A weak guide field points in the direction of the HFT, so that there is no null point (i.e., no fan-spine topology) at the location of the HFT. It is also noteworthy that the ratio of total flux in AR~11226 to the total flux in the two other ARs decreased considerably from June~3 through 7, due to particularly strong flux cancellation in AR~11226. Therefore the flux lobes of AR~11226 and 11227 are likely more symmetric at the time just before the eruption, i.e., somewhat closer to the simulation configuration (c.f. Figure  \ref{fig_qslcut})(b) and (c).

\begin{figure*}[t]
\centering
\includegraphics[width=1.\linewidth]{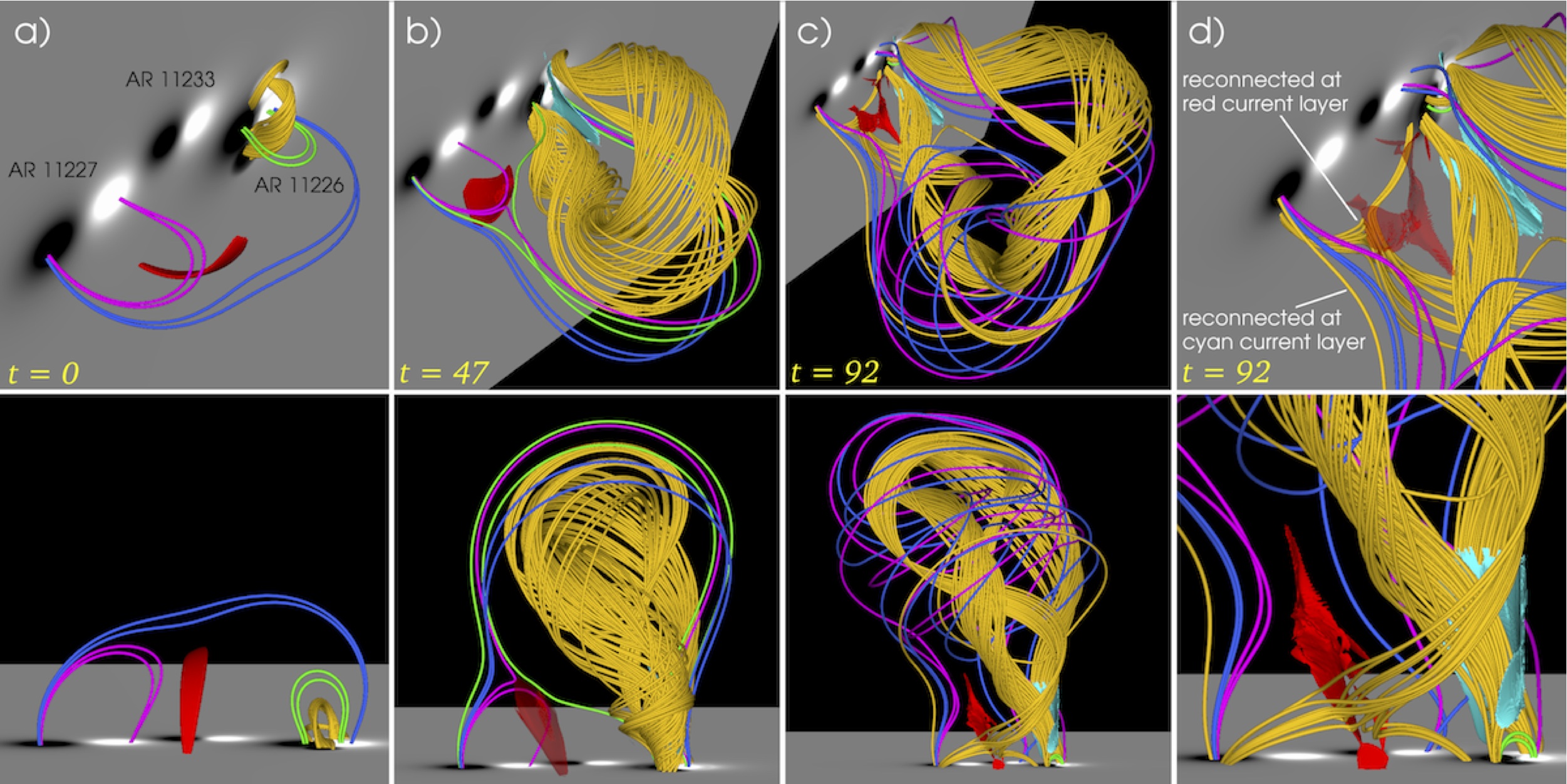}
\caption
{
Snapshots from the MHD simulation, showing a view similar to the SDO observations (top) and a side view on the modeled AR complex (bottom). The bottom plane shows $B_z > 0$ (white) and $B_z <0$ (black). The bipolar regions on the left, center, and right represent ARs 11227, 11233, and 11226, respectively. Selected fieldlines are shown to visualize reconnection between AR 11227 and AR 11226 during the eruption of the flux rope. 
Important current concentrations are shown by (opaque or transparent) red and cyan iso-surfaces of $j/B$ at varying values. 
(a) After the initial relaxation. An arched current layer (red) has formed at the HFT (cf. Figure\,\ref{fig_qslcut}(c)). 
(b) Expansion of the flux rope (golden fieldlines), triggering tilting and broadening of the red current layer and subsequent reconnection between potential fieldlines of AR 11227 (magenta) and AR 11226 (green). A second current layer (cyan) forms below the erupting flux rope. 
(c) Later state of the eruption. Further reconnection has occurred in both current layers. As a result, flux rope fieldlines are now connected to the negative polarity of AR 11227, and fieldlines rooted there (magenta and blue) have become part of the CME.
(d) Zoomed view into the reconnection regions at the same time as in panel c. 
}
\label{fig_simulation}
\end{figure*}

\subsection{MHD simulation}
\label{ss:MHD_results}
In support of our conjecture that the observed redirection of filament material and the associated coronal brightening were signatures of magnetic reconnection between the CME and AR 11227, we performed an MHD simulation as described in Sect.~\ref{sim_method}. The dynamic evolution of the simulated eruption is shown in Figure~\ref{fig_simulation}, and can be appreciated in the animation accompanying Figure~\ref{fig_summary} below, which contains co-temporal AIA 171~\AA\ and 304~\AA\ observations and corresponding simulation frames. The simulation reproduces to a very good approximation the huge expansion and global structure of the erupting flux rope, as well as the formation of a current layer consistent with the location of the coronal brightening. The formation of such a layer and subsequent reconnection are expected in a magnetic topology like the one studied here, if the system is subject to a sufficiently strong perturbation, as in a CME. In the following, we describe the main field line connectivity changes that occur as a result of reconnection in the simulation. 

Figure~\ref{fig_simulation}(a) shows the configuration after the initial relaxation, before the converging flows used to trigger the eruption are applied. The similarity of the field line connectivities with the extrapolated
fields shown in Figures~\ref{fig_qsl} and \ref{fig_qslcut} is obvious (see also Figure \ref{fig_qslcut}(c); the purple field lines shown in that figure are omitted here for clarity). The simulation field model also contains an HFT, running
predominantly along the y direction. As for the extrapolated fields, the presence of a weak guide
field precludes the presence of a null point in the HFT.  

The flux rope is visualized by two flux bundles (golden fieldlines), both of which are rooted in the rope's positive polarity footpoint. The outer, flat bundle is located at the rope's periphery, while the inner, arched bundle is located close to the flux rope center. The remaining fieldlines represent initially potential flux. An arched current layer (red iso-surface) has formed at the HFT (cf. Figure\,\ref{fig_qslcut}(c)), as a result of the dynamics present during the relaxation (see Figure~4 in \citealt{Torok11b} for a similar case). 
 
As the flux rope erupts and expands (Figure~\ref{fig_simulation}(b)), the green fieldlines located directly above it are pushed against the current layer from the top right direction, leading to a tilt of the layer that is favorable for reconnection between the green and magenta fieldlines. 
Once the current density has sufficiently steepened, these two flux systems indeed reconnect, producing new, low-lying connections between AR 11226 and 11227 (not shown in the figure, but see Figure~3 in \citealt{Torok11b} for a similar case); additionally, connections are formed that arch over the whole AR complex. However, this process is not expected to produce the observed redirection of filament material, since flux rope fieldlines have not yet reconnected. The eruption of the flux rope leads to the formation of a second, vertical current layer between the rope's legs (the canonical ``flare current sheet''; cyan iso-surface). 

Following the eruption further, connections between the flux rope and AR~11227 develop in two ways (Figures~\ref{fig_simulation}(c) and (d)). These correspond to the observation of brightenings on \textit{either} side of the QSL over the negative polarity in AR~11227, as shown in Figure~\ref{fig_qsl}. First, as the lower part of the expanding southern flux rope leg approaches the red current layer, field lines at its periphery are dragged into the reconnection region and reconnect with flux rooted in the negative polarity of AR~11227 (see the zoom in Figure~\ref{fig_simulation}(d)). The location of this current layer and the bottom part of the reconnected field lines correspond to the location of the EUV brightening at the reconnection region (Figures~\ref{fig_qsl}  and \ref{fig_summary}) and to the path of the redirected filament material, respectively, supporting our interpretation of the data. We can tentatively deduce from the simulations that the long (eastward-pointing) arm of the Y-shaped brightening corresponds to the bottom part of the reconnected fieldlines. 
The westward-pointing arm of the brightening corresponds to flux rope fieldlines that did not (yet) reconnect -- they represent the inflow region of the reconnection process. The short southwestward-pointing third arm illuminates the segments of these two sets of fieldlines above the reconnection region, which lie next to each other in the downward-pointing flux of the southern flux rope leg. The remaining element of the reconnection process is flux originating in the positive polarity of AR~11227 (similar to the magenta fieldlines in panel (a) before reconnection), connecting to the negative polarity of AR~11226 after reconnection. These do not carry filament material and are not included in the figure, to avoid a crowded appearance. In this interpretation, the current sheet is located between the leftward and rightward-pointing arms of the brightening and passing through the short third arm. 

Second, field lines initially or temporarily arching over the expanding flux rope reconnect in the cyan current layer with flux rope field lines and become part of the CME. The latter include blue field lines and magenta field lines that have previously reconnected at the red current layer and temporarily overarch the flux rope as well. Both reconnection processes occur more or less simultaneously and result in new connections between the erupting flux rope and the negative polarity of AR~11227. However, the reconnection occurring at the cyan current layer cannot be expected to be responsible for the observations of the deviating filament material, since (i) it will not produce a brightening at the observed location in the corona (Figure \ref{fig_summary}(a--c)), which we associate with the red current sheet (Figure \ref{fig_simulation}(d)) and (ii) the resulting fieldline connections between AR~11226 and AR~11227 are made up for most of their length from initially potential field that cannot carry filament material. 

In order to clarify the role of the large lateral expansion of the erupting filament and investigate whether or not it is indeed the driver of the reconnection between the erupting flux rope and field lines of AR~11227, we computed the maximum current density in the red current layer for the times shown in Figure~\ref{fig_simulation}(a--d), and found a strong (15-20-fold) enhancement of $J$, before reconnection begins in this (red) current sheet in the simulation.  Reconnection is much more likely to occur, or to become faster, when the current sheet is getting thinner and, correspondingly, $J$ is becoming higher. \cite{Titov_etal03} and \cite{Galsgaard_etal03} found that an HFT can be pinched into a current sheet in the presence of appropriate external driving motions. In this case, it is most likely that such driving motions were mainly provided by the lateral expansion of the unstable flux of AR 11226. Therefore, we conclude that the lateral expansion of the flux rope erupting in AR~11226 was the major driver of its interaction/reconnection with AR~11227.  

We note that the overall reconnection in the simulation is more complex than just described. For example, a significant number of fieldlines reconnect subsequently in both current layers, and some appear to do so several times, repeatedly changing their connectivity. Also, there is reconnection occurring at a current layer that forms between AR~11233 and AR~11226 (not shown in Figure~\ref{fig_simulation} for clarity). As a consequence, magenta and blue field lines, starting in AR~11227 and arching over the erupting flux, change their footpoints a second time to connect to AR~11233 (Figure~\ref{fig_simulation}(d)). A detailed investigation of the evolution and interplay of all reconnection processes is beyond the scope of this paper. The reconnection primarily relevant in the present context is the one occurring in the red current layer between the southern flux rope leg and the flux initially connecting the polarities within AR~11227, as described above.

\begin{figure}[htpb]
\centering
\includegraphics[width=1.\linewidth]{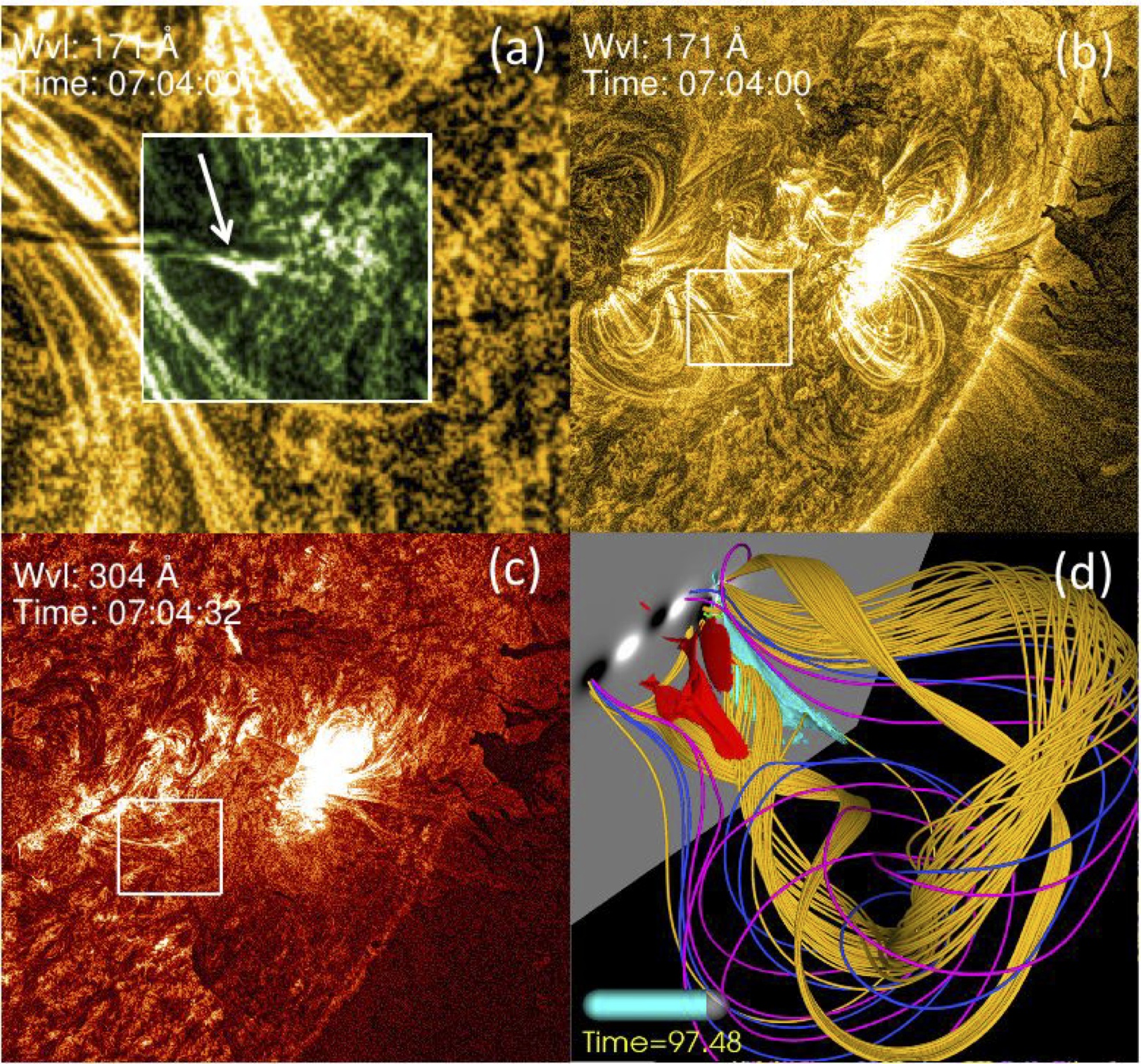}
\caption
{
A representative frame from the accompanying summary movie (movie3.mov), showing {\sl SDO}/AIA images and the MHD simulation of reconnection between topologically distinct domains: filament eruption/CME and a neighboring AR.   Images in the AIA 171 \AA~ 
channel show the reconnection region in the solar corona in a zoomed-in image (a) indicated by a white rectangle in panel (b). The reconnection region (highlighted in green and indicated by an arrow) is located at the interface of the magnetic flux rope that contains the erupting filament and magnetic loops of the neighboring AR 11227. New connections channel filament material towards the magnetic footpoints of the neighboring AR. As reconnection proceeds, coronal plasma undergoes direct heating in the unusually high-density plasma environment due to the presence of cool and dense filament material. Image from the 304 \AA~ 
channel (c) shows the dense absorbing plasma of the erupting filament over the AR and falling filament material (above the solar limb). 
MHD simulation (d) at a somewhat later time than in Figure\,\ref{fig_simulation}(c), showing a thicker flux rope core (golden fieldlines), fieldlines originally rooted in AR 11227 in magenta, and the two main current layers for a constant value of $j/B$ in red and cyan. \\ 
(A color version and an animation of this figure (movie3.mov) are available in the online journal.)
}
\label{fig_summary}
\end{figure}

\section{Discussion and conclusions}
\label{conclusion}
We present combined \textsl{SDO}/AIA observations, magnetic extrapolation and QSL computation based on \textsl{SDO}/HMI magnetic data, and a data-constrained MHD simulation of the 2011 June 7 massive filament eruption, providing the first direct evidence of large-scale magnetic restructuring via reconnection during the CME eruption process. The eruption originated from AR 11226, the westernmost in a three-AR complex (Figure \ref{fig_config}).
As the massive filament was rising in the solar atmosphere, the underlying magnetic flux rope underwent fast lateral expansion, triggering an interaction with the flux of the neighboring AR~11227 (Figures \ref{fig_304}, \ref{fig_simulation}, \ref{fig_summary}, and the accompanying animations movie1.mov and movie3.mov).
This was observed as a deviation of filament material flowing back to the surface along the SE leg of the expanded flux rope. Part of the material was redirected along a path deviating sharply from the flux rope leg and pointing into AR~11227, while the remaining material continued to flow back into AR~11226. This redirection represents unambiguous evidence of reconnection, since down-flowing filament plasma is guided by the magnetic field in a low-plasma-beta ($\beta<1$) environment. While the particularly massive NW leg of the erupted filament did show a violation of this standard picture in form of plasma blobs indicating the RT instability,  the SE leg was less prone to that. Moreover, since the new direction of motion was neither set by the inertia of the down-flowing plasma, nor by gravity, it must have been set by the magnetic field. Further support for the interpretation of the deviation as resulting from reconnection is given by the brightening of the filament material in the vicinity of the point of deviation, by the existence of an HFT in the QSL map at this location, and by the occurrence of reconnection of the inferred topology at the observed location in our MHD model of the eruption. The location of the current sheet and the new connection correspond very well to the location of the observed Y-shaped brightening and the path of the down-flowing filament material that was redirected towards AR~11227. The occurrence of the brightening was supported by the unusually high density of the back-flowing filament material; this allowed the first imaging of the exact location of a coronal reconnection region.

The density of the filament material in the vicinity of the reconnection region was estimated in \citet{Williams13} to be $\sim10^{10}$~cm$^{-3}$, more than one order of magnitude higher than the typical density of bright coronal loops. We conjecture that this facilitated the local heating underlying the brightening by providing sufficient density so that particles, accelerated in the inferred reconnection region, are thermalized by collisions locally in the corona, a process normally efficient only when the particles precipitate to the chromosphere. It is likely that this circumstance made the present event special, since the also expected heating by the slow-mode reconnection shocks does not normally ``illuminate'' the vicinity of a reconnection region in the corona.
The most prominent brightening at the reconnection region is seen in the AIA 171 \AA~ waveband, indicating that the filament plasma was heated by more than one order of magnitude to temperatures $\gtrsim 6 \times 10^{5}$ K. 

A data-constrained MHD simulation of the event demonstrates that the lateral expansion of the rising flux led to flux pile-up and current sheet formation at an HFT between AR~11226 and 11227, which is also revealed by the QSL computation. The flux rope expansion was found to steepen the current density in the (red) current layer by a factor of about 15--20 before reconnection sets in, which clearly indicates the role of lateral expansion as the driver.  Reconnection in this current sheet formed a new connection between the expanding CME flux rope rooted in AR~11226 and the flux rooted in AR~11227. 

In summary, in this study we presented clear evidence of an erupting flux rope interacting with the surrounding coronal magnetic field via magnetic reconnection. Our topological analysis of the three-AR complex identified an HFT, a preferential location for magnetic reconnection. Our data-constrained MHD simulation has showed reconnection indeed taking place there in a way consistent with the observations. Finally, for the first time, we presented observation of plasma being heated in situ in the corona at a location consistent with the site of reconnection region predicted by the models.  

Another imaging evidence of coronal magnetic reconnection was recently presented by \cite{Su13}, for a confined limb flare on 2011 August 17. These authors found horizontally approaching cool and hot loops and newly formed V-shaped hot loops leaving a hot ($> 10$ MK) vertical sheet, the reconnection region, in the upward and downward directions. The perspective of that observation pointed along the current sheet (the flare current sheet in their case); thus it allowed all four sections of reconnecting fieldlines, before and after their reconnection, to be directly imaged as coronal loops. The Y-shaped brightening in the event investigated here, located at the upper end of a new connection to a neigboring AR, also allows one to infer the occurrence of magnetic reconnection unambiguously, although not all four fieldline sections involved are illuminated by hot plasma. The downflow of filament plasma into the reconnection region along one arm of the brightening and out of the reconnection region along the other two arms suggests the illumination of non-reconnected and reconnected fieldlines in a manner indicated by the QSL computation and confirmed by the MHD simulation. 

In the SDO movies of various wavelengths and temperatures we see and can say with confidence that magnetic reconnection was taking place between the erupting and neighbouring magnetic structures. However, when focussing on  the bright reconnection region, it is difficult to determine small-scale details of the reconnection process. One of the reasons is that a sequence of complex 3D structural changes were captured in 2D images. There were a multitude of bipoles within the three ARs which determined the presence of the HFT but also made the overall topological structure complex. 
Another reason is that magnetic reconnection at the coronal level is only explicitly imaged when dense plasma blobs are passing at the right location and time.  Magnetic reconnection, as manifested by heating, was clearly taking place at several locations, not only at the most prominent site discussed in the article so far. Reconnection was likely to occur also  at places which escaped to be observed.  Some details of the observed event could also be compatible with secondary reconnections involving other connectivity domains. Similar complexity is present in the simulation as described at the end of Section \ref{ss:MHD_results}. 

 
The magnetic reconnection process presented here involved only a tiny fraction of the flux contained in the hugely inflated erupting flux rope, leading to only weak energy release.
However,  this low-energy reconnection took place in an unusually high-density plasma environment created by the down-flowing dense filament material. This enabled us to provide imaging evidence of a physical process that under normal circumstances does not have observable signatures even with current state-of-the-art instrumentation. 

These results provide the first direct observational evidence that CMEs, while expanding into a magnetically structured corona, reconnect with surrounding magnetic structures, leading to a large-scale re-configuration of the coronal magnetic field. When the expanding flux rope of a CME encounters external QSLs or separators, like in the present case, its forcing may ``activate'' the pre-existing topological structure, triggering reconnection, leading to immediate restructuring, as in the case presented here. In addition,  sympathetic events may occur with some delay after the large-scale field has been modified in this way \citep{Schrijver11,Torok11b}. 

\acknowledgments{The authors thank the {\sl SDO}/AIA and HMI consortia for the data, and the use of jHelioviewer (http://jhelioviewer.org/) for browsing the data and making movie1.mp4, which is presented in the electronic edition of this article.  We are grateful to Huw Morgan for the use of his superb image processing method to produce some of the figures and movies presented in this article. The research leading to these results has received funding from the European Union's Seventh Programme for Research, Technological Development and Demonstration under Grant Agreement No. 284461 (eHEROES project). LvDG's work was supported by the Hungarian Research grants OTKA K-081421 and K-109276. LvDG and DB acknowledge support by STFC Consolidated Grant ST/H00260/1. TT was supported by NASA's HTP, LWS, and SR\&T programs, and by NSF through grant AGS-1249270. LMG is grateful to the Royal Society for a University Research Fellowship. JC thanks UCL and MPI for an Impact PhD Studentship.  BK acknowledges support by the DFG and by the Chinese Academy of Sciences under Grant 2012T1J0017.}

\bibliographystyle{apj}
\bibliography{Reconnection_7June11_v4b}  

\end{document}